\documentclass[manuscript]{aastex}
\usepackage{fancyhdr}
\usepackage{longtable}
\usepackage{latexsym}
\usepackage{graphicx}
\usepackage{amsmath}
\usepackage{natbib}
\usepackage{upgreek}
\usepackage{multirow}
\usepackage{array}
\usepackage{pdflscape}
\usepackage{graphicx,subfigure}
\usepackage{tabularx}
\usepackage{epsfig}
\usepackage{color}
\usepackage{lscape}
\usepackage{ulem}
\usepackage{amsmath,bm}
\usepackage{algorithmicx,algorithm}

\newcommand{\PreserveBackslash}[1]{\let\temp=\\#1\let\\=\temp}
\newcolumntype{C}[1]{>{\PreserveBackslash\centering}p{#1}}
\newcolumntype{R}[1]{>{\PreserveBackslash\raggedleft}p{#1}}
\newcolumntype{L}[1]{>{\PreserveBackslash\raggedright}p{#1}}

\shorttitle{Imbalanced learning for RR Lyrae stars}
\shortauthors{J. Zhang, Y. Zhnang \& Y. Zhao}

\begin{document}

\title{Imbalanced learning for RR Lyrae stars based on SDSS and GALEX databases}

\author{Jingyi Zhang \altaffilmark{1,2}, Yanxia Zhang\altaffilmark{1},
       Yongheng Zhao \altaffilmark{1}
}

\altaffiltext{1}{Key Laboratory of Optical Astronomy, National Astronomical Observatories, Chinese Academy of Sciences, Beijing 100012, China. \sf zyx@bao.ac.cn}
\altaffiltext{2}{University of Chinese Academy of Sciences, Beijing 100049, China.}

\begin{abstract}
We apply machine learning and Convex-Hull algorithms to separate RR Lyrae stars from other stars, like main sequence stars, white dwarf stars, carbon stars, CVs and carbon-lines stars, based on the Sloan Digital Sky Survey (SDSS) and Galaxy Evolution Explorer (GALEX). In the low-dimensional space, the Convex-Hull algorithm is applied to select RR Lyrae stars. Given different input patterns of ($u-g$, $g-r$), ($g-r$, $r-i$), ($r-i$, $i-z$), ($u-g$, $g-r$, $r-i$), ($g-r$, $r-i$, $i-z$), ($u-g$, $g-r$, $i-z$) and ($u-g$, $r-i$, $i-z$), different convex hulls can be built for RR Lyrae stars. Comparing the performance of different input patterns, $u-g$, $g-r$, $i-z$ is the best input pattern. For this input pattern, the efficiency (the fraction of true RR Lyrae stars in the predicted RR Lyrae sample) is 4.2\% with a completeness (the fraction of recovered RR Lyrae stars in the whole RR Lyrae sample) of 100\%, increases to 9.9\% with 97\% completeness and to 16.1\% with 53\% completeness by removing some outliers. In the high-dimensional space, machine learning algorithms are used with input patterns ($u-g$, $g-r$, $r-i$, $i-z$), ($u-g$, $g-r$, $r-i$, $i-z$, $r$), ($NUV-u$, $u-g$, $g-r$, $r-i$, $i-z$) and ($NUV-u$, $u-g$, $g-r$, $r-i$, $i-z$, $r$). RR Lyrae stars, which belong to the class of interest in our paper, are rare compared to other stars. For the highly imbalanced data, cost-sensitive Support Vector Machine (SVM), cost-sensitive Random Forest and Fast Boxes are used. The results show that information from GALEX is helpful for identifying RR Lyrae stars and Fast Boxes are the best performers on the skewed data in our case.

\end{abstract}


\section{Introduction} \label{sec:intro}

With the rapid development of astronomical observation technologies, various large survey projects of different wavelengths are finished, on-going and in plan, such as 2MASS \citep{sk06}, SDSS \citep{york00}, GALEX \citep{bh11}, WISE \citep{wr10}, LSST \citep{iz08} and others. The observed data from different bands have grown dramatically, which promotes the formation of multi-band astronomy. The observation and research on the multiwavelength nature of celestial objects have been opened up. These multiwavelength data are of great value for the study of objects. Faced with so massive amount of astronomical data, we should extract the useful information from the data by automatic methods.

As a tracer, RR Lyrae stars can be used to study the structure and evolution of the Galactic halo \citep{bu01}. Their absolute magnitude is nearly a constant value, which is better for calculating the distances and refining models to measure the Galaxy's mass as the standard candles \citep{kk01}. So it is important to pick out RR Lyrae stars from large survey data. Many works have been performed by the selection criteria in the two-dimensional color space. \citet{dj14} used the $Ks-W3$ and $g-i$ color space and selected 199 infrared excess candidates. \citet{fi00} applied the $r-Ks$ and $J-Ks$ color-color diagrams to trace the stellar color locus and classified the main sequence stars. But in a two-dimensional color space, RR Lyrae stars are not easy to be separated better from other sources, e.g. Cepheid variables, quasars, normal spectral class A, F and M stars. Nevertheless \citet{fan99} presented the stellar distribution in the simulated color space of ($u-g, g-r, r-i$) and got an intuitive understanding of the stellar distribution. \citet{kk98} focused on the identification of the RR Lyrae stars based on the SDSS colors. \citet{iz00} selected 148 RR Lyrae star candidates from SDSS commissioning data for about 100 deg$^2$ of sky. \citet{iz03} also picked out RR Lyrae stars with SDSS DR1 and found that part of the candidates were the remnant of the Sagittarius dwarf galaxy disrupted by the tidal force of the Milky Way. Apparently, it is possible to recognize RR Lyrae stars only from their single-epoch data.

Besides the simple color selection criteria, machine learning algorithms are applied to select RR Lyrae stars. Random Forest is of importance to detect members of rare classes. Many studies for selecting RR Lyrae stars are based on the Random Forest algorithm \citep{gc14,hn16,ad16}. \citet{sb17} explored XGBoost to identify RR Lyrae stars from the PS1 sample with the fitting light curves. To construct an automated approach for targeting RR Lyrae stars, \citet{ef16} constructed a good performer based on the AdaBoost family of classifiers.

In fact, the rare or unknown cases are our interest in astronomy, e.g. identifying these rare objects in large survey databases, separating RR Lyrae stars from quasars, detecting gravitational lenses from image databases. RR Lyrae stars belong to this class and are the minority compared to the whole star sample. As a result, there is a thorny problem, which is the high imbalance of data. In general, classical machine learning algorithms do not work well with imbalanced datasets, mainly because they attempt to reduce the overall misclassification error which biases the classifier towards the majority class in imbalanced datasets. In order to select the minority, main approaches are raised to deal with class imbalance, including re-sampling or re-weighting, changing search strategies in learning, cost-sensitive learning, one-class learning, using another measures, adjusting classification strategies, using hybrid and combined approaches (boosting like re-weighing) and others. Sampling methods include oversampling, undersampling and synthesizing new minority classes, like SMOTE \citep{ch02}. The cost-sensitive learning methods are hot in machine learning, and aim to get a minimal cost result on classifying the skewed data \citep{wg07}. Cost-sensitive learning makes the original machine learning algorithm adapt to the needs of the real application to some extent, by the cost imbalance of different miscarriage cases in the real world. Support Vector Machine (SVM) with cost-sensitivity, Random Forest with cost-sensitivity and Fast Boxes were applied to solve the imbalance of classes \citep{rc14}. Their results showed that Fast Boxes were better for separating the interesting class from other classes accounting for the majority.

In this work, our primary goal is to separate the RR Lyrae stars from the other types of stars. This paper is organized as follows. Section~2 provides a brief description of the SDSS and GALEX surveys. How to get the samples is also presented in detail. In Section~3, we describe the algorithms and apply them on the samples. Section~4 discusses the performance of the approaches in our case and analyses the results. Finally, we give the conclusions and the future work.

\section{The data} \label{sec:data}

SDSS \citep{york00} is a major multi-filter imaging and spectroscopic redshift survey using a dedicated 2.5-m optical telescope, which provides a wide field of view, and has five filters located in the $u$, $g$, $r$, $i$, and $z$ bands, which span wavelengths from 0.36 to 0.90 $\mu$m. The processed data include all photometric and spectroscopic observations. Compared to the prior SDSS Data Release, SDSS Data Release 13 (DR13), the first release of the Sloan Digital Sky Survey IV, contains new reduction and improved spectral calibration. DR13 is cumulative and provides more robust and precise photometric data.

GALEX \citep{bh11} is a space telescope, launched by a Pegasus rocket into the orbit in 2003. It covers wavelengths from 135 to 280 nm. Near- and far-UV emissions are measured by it.

Firstly, a sample of stars with spectral classification is produced. To create a reliable and complete sample of stars, we need to limit the original SDSS data. Using the DR13 SkyServer and access to the Catalog Archive Server (CAS) database, we extract the point sources with their spectral classification. They are collected from the SpecPhotoAll table of SDSS DR13. Some sources should be removed because either observed multiple times or have spectroscopic data but lack photometry. In our work, we set the class = `Star' to select the stellar objects. To eliminate the sources observed under bad conditions, we set the SDSS flags with BRIGHT=0, SATURATED=0, EDGE=0, NOPROFILE=0 and BLENDED=0, as provided by \citet{dj14}. In addition, the PSF magnitudes used are de-reddened and corrected for Galactic extinction by the dust maps of \citet{sc98}. Imposing sciencePrimary =1 and Mode = 1, we select the sources with best spectra and best photometries. Due to the impact of magnitude errors, there is a large deviation in the distribution of stars in the color-color diagram. In the $u$, $g$, $r$, $i$, and $z$ bands, the resolution of $u$ band and $z$ band is lower than $g$, $r$ and $i$ bands, so we set the psfMagErr$\_u<$0.2, p.psfMagErr$\_g<$0.1, p.psfMagErr$\_r<$0.1 and p.psfMagErr$\_i<$0.1, p.psfMagErr$\_z<$0.2. Finally 511,201 sources are obtained, which consist of main sequence stars, white dwarf stars, carbon stars, CVs and carbon-lines stars.

The sample of RR Lyrae stars are got from the SDSS Stripe 82. There are 483 RR Lyrae stars \citep{sb07}. The region on the celestial equator, dubbed ``Stripe 82", is imaged in more than 10 times and gives co-added optical data 2 times deeper than single epoch SDSS observations. It is superior in studying many time-varying objects, such as variable stars and quasars. The repeated observations also make the photometric data of non-varying stars more reliable.

To identify wether RR Lyrae stars can be separated from other types of stars towards the blue end of light, cross-match of 511,684 sources (483 RR Lyrae stars and 511,201 others) with the GALEX database is performed within a radius of 3.0 arcsec. Finally 169,719 sources are selected and machine learning algorithms are applied to them.

\section{The method} \label{sec:method}

\subsection{Convex hull algorithm} \label{subsec:convex}

Convex hull is a concept in computational geometry. In a real vector space $V$, the intersection $S$ of all convex sets $X$ for a given convex set $X$ is called the convex hull of $X$. The convex hull of $X$ can be constructed by a linear combination of all points ( $X_1$, $X_2$, $\cdots$, $X_n$ ) in $X$, which can be imagined as a rubber band just wrapped in all points in the two-dimensional Euclidean space. For more information of convex hull, please refer to \citet{sk82}. Given a set of points on a two-dimensional plane, the convex hull is a convex polygon formed by concatenating the outermost points, which contain all the points in the data set. Convex hull can be constructed for the sample using Qhull\footnote{www.qhull.org}, which computes the convex hull for the 2-d, 3-d, 4-d and higher dimensional data. To get a more visual result, we only compute the convex hull in the 2\-d and 3\-d spaces.

\subsection{SVM} \label{subsec: svm}

Support vector machine (SVM), proposed by \citet{vv95}, is a supervised learning algorithm for classification and regression analysis, which separates the points using a hyperplane. If the hyperplane has the largest distance to the nearest training points of any class, a good separation can be obtained. The superiority of SVM is obvious for 2-class classification. Suppose that classifying $n$ points in real space $\mathbb{R}^m$, denoted by the matrix $X_{m\times{n}}$. SVM predicts the outputs $y_i=\pm 1$ of input vector $\bm{x_i} = (x_{1i}, x_{2i}, \cdots, x_{mi}, 1)^T, i = 1, 2, \cdots, n$. Finding the optimal hyperplane with a maximum margin by minimizing the following decision function, the training data may be separated by the hyperplane.

\begin{equation}
f(\bm{x_i}) = \sum_{i=1}^n l(y_i, \langle \boldsymbol{\omega}^T,\bm{x_i}\rangle) + C\Vert \boldsymbol{\omega} \Vert^2,
\end{equation}
where $l (y_i, \langle \boldsymbol{\omega}^T,\bm{x_i}\rangle) = \mathrm{max} (0,1-y_i \langle \boldsymbol{\omega}^T,\bm{x_i}\rangle)$ represents the loss function, $\boldsymbol{\omega}$ \textbf{is called the normal vector of the hyperplane} and $C$ denotes the penalty parameter.

There are many applications of SVM in astronomy. For example, \citet{pn13} made a combined classifier of SVM and KNN to select the quasar candidates. \citet{gd09} compared the performance of SVM with K-Dimensional Tree (KD-Tree), and got a better result to separate quasars from stars. In this work, input vectors are mapped into a high-dimensional space to separate RR Lyrae stars from other types of stars.

\subsection{Random Forest} \label{subsec: random forest}

Random Forest grows many decision trees to classify the data \citep{br01}. Given a new input vector, each tree votes for the class. Over all trees, the forest makes a choice and selects the classification having the most votes. The advantages of Random Forest are as follows \citep{gd09}:

\begin{enumerate}
\item Among many machine learning algorithms, the accuracy of Random Forest is high.
\item It can handle a very large number of input variables efficiently.
\item It estimates which variables are important in the classification problems.
\item In the forest building progress, it produces an inherent unbiased estimate of the generalization error.
\item It has an efficient way of estimating missing data and remains accuracy when a large fraction of the data are missing.
\item It can balance errors in unbalanced data sets.
\item It calculates the approximations between cases that can be useful for clustering, outlier detection and giving interesting data views.
\end{enumerate}

\subsection{SVM and Random Forest with cost-sensitivity} \label{subsec: cost-sensitive learning}

Although SVM and Random Forest are popular machine learning algorithms owing to their practical advantages, the traditional SVM and Random Forest are not applicable to imbalanced data. Usually for the balanced data, the class weight of the positive and the negative class is same. But in this work, we are more interested in the minority. If not altering the distribution of skewed data, the prediction is inclined to the majority. The most approaches are to directly use a learning algorithm ensembling cost-sensitivity. Namely, a learning algorithm factors in the costs when building a classification model. For SVM and Random Forest, we all set the parameter $C$ of class $i$ to class\_weight $[i]*C$ for adjusting the imbalance of cost, called cost-sensitive learning. So far there have been many works focusing on this aspect, for example, \citet{ya17} handled the imbalance of data set with a cost-sensitive approach. For each of considered classes, SVM is modified to incorporate a varying penalty parameter $C$. \citet{yi14} proposed a cost-sensitive algorithm based on Random Forest. The cost-sensitive Random Forest balanced the data by bagging and altered the misclassification cost, and their results showed that the cost-sensitive Random Forest achieved higher performance. Here the programs of SVM and Random Forest with cost-sensitivity are adopted from scikit-learn \citep{pe11}.

\subsection{Fast Boxes} \label{subsec: cost-sensitive learning}

The recent Fast Boxes algorithm is also used to compare the performance with the baseline algorithms. Fast Boxes are also a cost-sensitive learning algorithm, where relative costs are implicitly determined during training. Data are assumed to lie in a grid. The minority class is clustered and a minimal enclosing box in the grid is determined around each cluster. The feature space is partitioned based on the boxes and finally the boundaries are expanded using an exponential loss function \citep{sh15}. More details about the algorithm of Fast Boxes and three stages of Fast Boxes are as follows \citep{rc14}:

\begin{enumerate}
\item Characterizing: the positive class is clustered into $K$ clusters, which is a one-class learning method and sets decision boundaries around the clusters.
\item Dividing space: the space is divided into subspaces according to the boundaries of the negative and the positive class.
\item Boundary expansion: using a regularized exponential loss, the boundary is extended to contain more positive class.
\end{enumerate}

\begin{algorithm}
\caption{The Fast Boxes algorithm}
\hspace*{0.02in} {\bf Input:}
the number of clusters $K$, weight $c$, expansion parameter and data\\
\hspace*{0.02in} {\bf Output:}
boundaries of boxes
\begin{algorithmic}[1]
\State Scale the features (different input patterns) to be between -1 and 1.
\State Cluster the RR Lyrae stars into $K$ clusters.
\State Compute the starting boundaries $l_{s,j,k}$ and $u_{s,j,k}$ for each cluster, the $j$-th dimension lower and upper boundaries for the $k$-th cluster.
\State Construct local classifiers $X_{l,j,k}$ and $X_{u,j,k}$.
\State For avoiding numerical problems, multiply each term by $\exp (1)$ and revise the lower and upper boundaries.
\State Perform expansion to guarantee the box is always expanding,  which implies that this algorithm includes as many minority classes of samples as possible.
\State Un-normalize by rescaling the input patterns back to get meaningful values.
\State Counts of true positive and false positive class members.
\end{algorithmic}
\end{algorithm}

\subsection{Evaluation Metric} \label{sec: metric}

In a low-dimensional space, completeness (Recall) and efficiency (Precision) are used to evaluate the convex hull algorithm. The Precision for a class is the number of true positives divided by the total number of elements labeled as belonging to the positive class; Recall is defined as the number of true positives divided by the total number of elements that actually belong to the positive class. In our case, the completeness is referred as the fraction of recovered RR Lyrae stars in the whole RR Lyrae sample and the efficiency is defined as the fraction of true RR Lyrae stars in the predicted RR Lyrae sample \citep{sb07}. Often, there is an inverse relationship between Precision and Recall, where one increases at the cost of reducing the other. In general, we can not discuss Precision and Recall scores in isolation. Instead, one of these two measures is given at a fixed level of the other measure or both are combined into a single measure, such as the F-measure. Precision-Recall (PR) curve is the plot of Recall on the x-axis and Precision on the y-axis (i.e. the trade-off between efficiency and completeness), which gives an informative picture of an algorithm¡¯s performance especially for highly skewed datasets. A high area under the curve in PR space means both high recall and high precision, where high precision corresponds to a low false positive rate, and high recall refers to a low false negative rate. High scores for both measures indicate that the classifier is giving accurate results (high precision), as well as giving a majority of all positive results (high recall).

In a high-dimensional space, the different cost ratios (the cost for the majorities) are used to evaluate the performance of machine learning algorithms. For every algorithm, each weight corresponds to a point on the receiver operating characteristic (ROC) curve, which is a graphical plot that illustrates the diagnostic ability of a binary classifier system as its discrimination threshold is varied. ROC analysis provides tools to select possibly optimal models and to discard suboptimal ones independently from (and prior to specifying) the cost context or the class distribution. Actually, the cost information used in the paper is unknown and inferred based on the presumption. Comparing the different methods, we choose the area under the convex hull of the ROC curve (AUH) \citep{pfft01}, which is often used as an evaluation metric to select the best performer. The counts of true positive as a function of the counts of false positive is presented on the ROC curve. Finally the AUH is normalized as usual by dividing it by the number of positive examples times the number of negative examples. The top left corner of the plot is the ``ideal" point and the ideal best AUH is 1.

\section{Performance of the algorithms} \label{sec:performance}

Our aim is to pick out RR Lyrae stars from the whole star sample. The performance to select RR Lyrae stars is influenced by some factors, such as sample imbalance, the performance of a classifier. The number of known RR Lyrae stars is 483 in our case. Therefore selecting RR Lyrae stars from other stars belongs to heavy imbalance problem. Here we plan to explore convex hull to sort RR Lyrae stars in the low-dimensional spaces, and apply cost-sensitive SVM, cost-sensitive Random Forest as well as Fast Boxes to choose RR Lyrae stars in the high-dimensional spaces.

\begin{figure}
\centering
\includegraphics[bb=288 18 940 660,width=15cm]{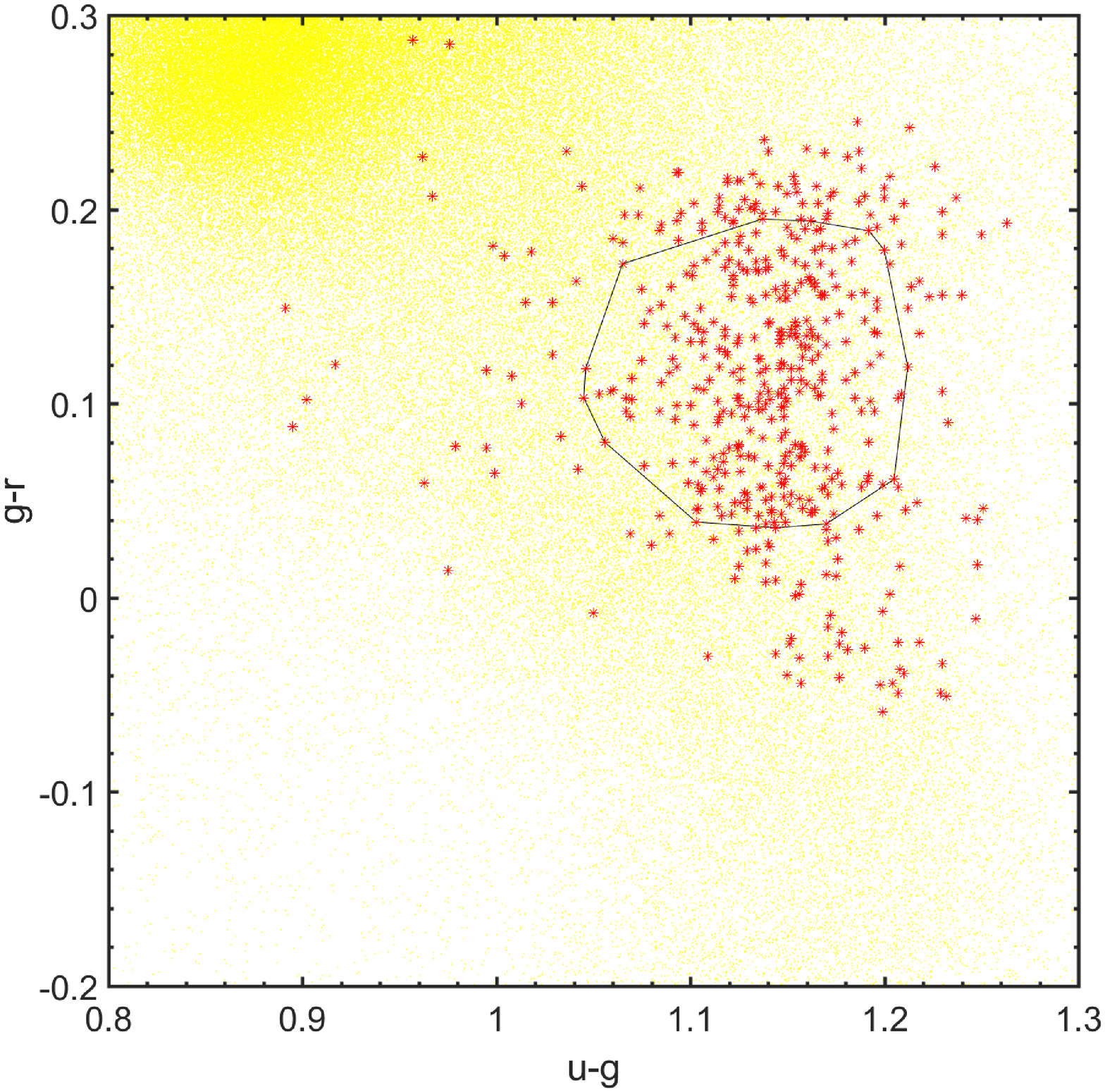}
\caption{Convex hull built in the $u-g$ and $g-r$ color space at 53\% completeness\label{fig:Fig1}. The red asterisks represent RR Lyrae stars, and the yellow dots represent other stars.}
\label{fig1}
\end{figure}

\begin{figure}
\centering
\includegraphics[bb=288 18 940 660,width=15cm]{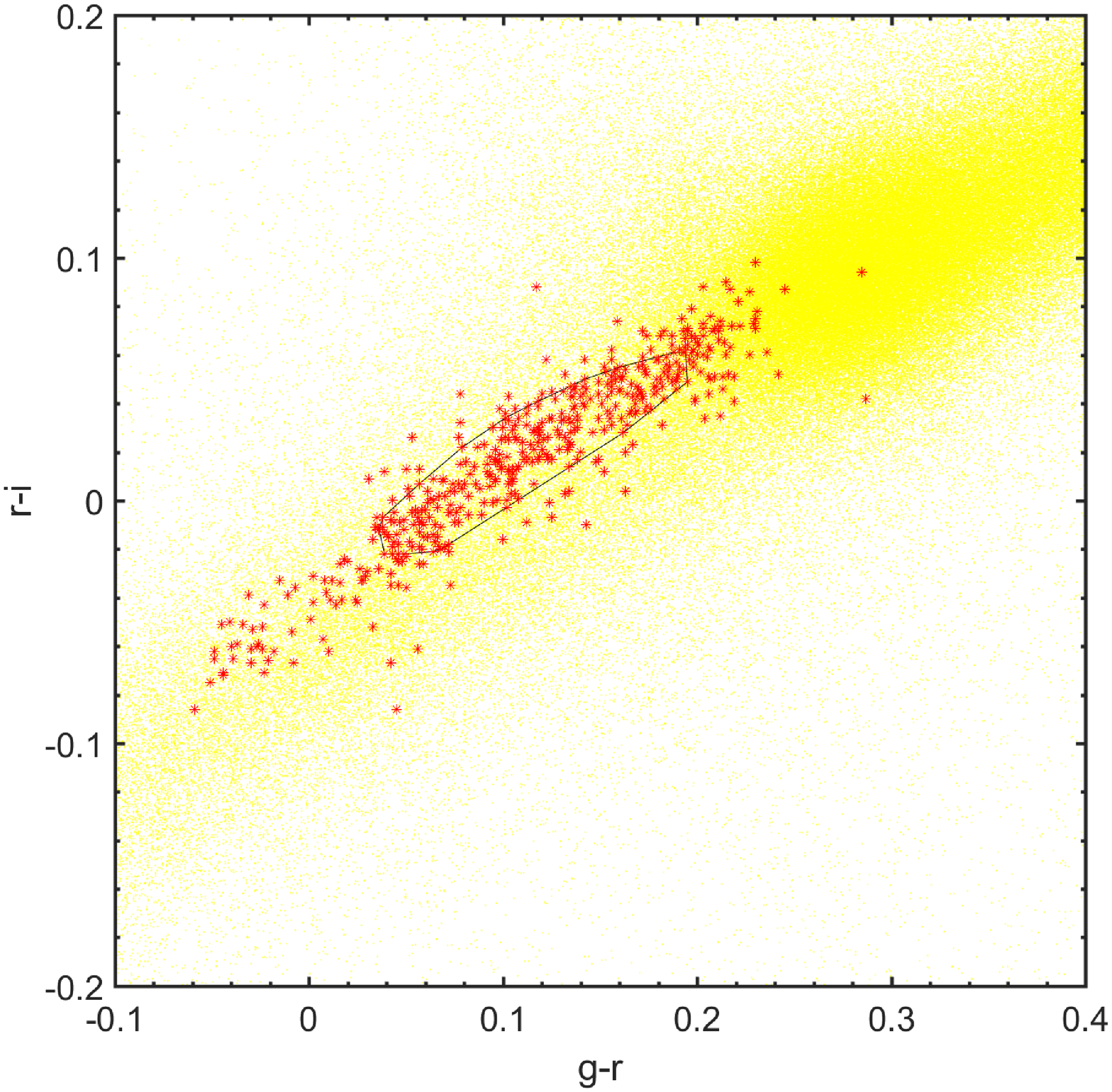}
\caption{Convex hull built in the $g-r$ and $r-i$ color space at 53\% completeness\label{fig:Fig2}. The red asterisks represent RR Lyrae stars, and the yellow dots represent other stars.}
\label{fig2}
\end{figure}

\begin{figure}
\centering
\includegraphics[bb=288 18 940 660,width=15cm]{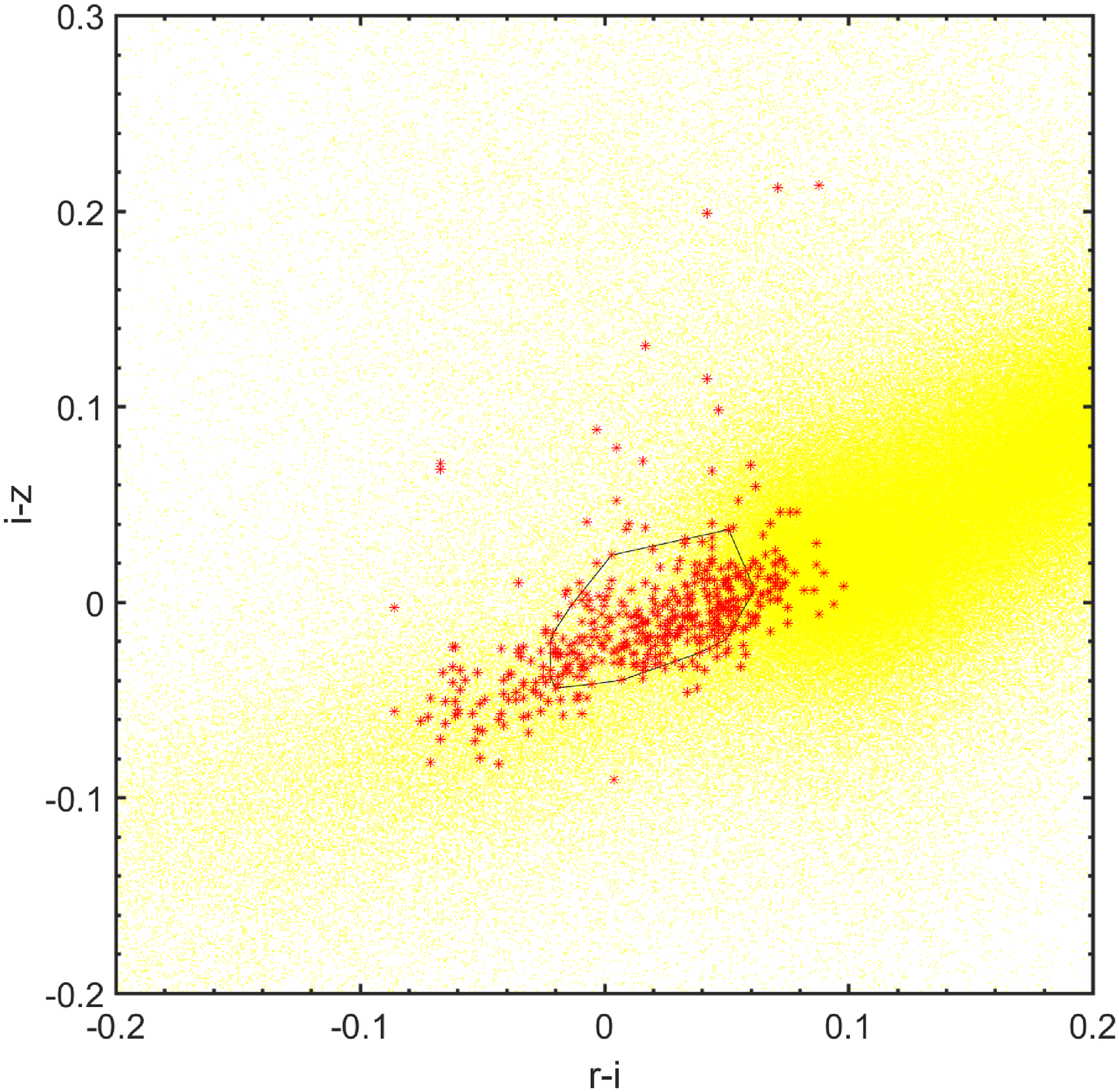}
\caption{Convex hull built in the $r-i$ and $i-z$ color space at 53\% completeness\label{fig:Fig3}. The red asterisks represent RR Lyrae stars, and the yellow dots represent other stars.}
\label{fig3}
\end{figure}

\begin{figure}
\centering
\includegraphics[bb=288 18 940 660,width=15cm]{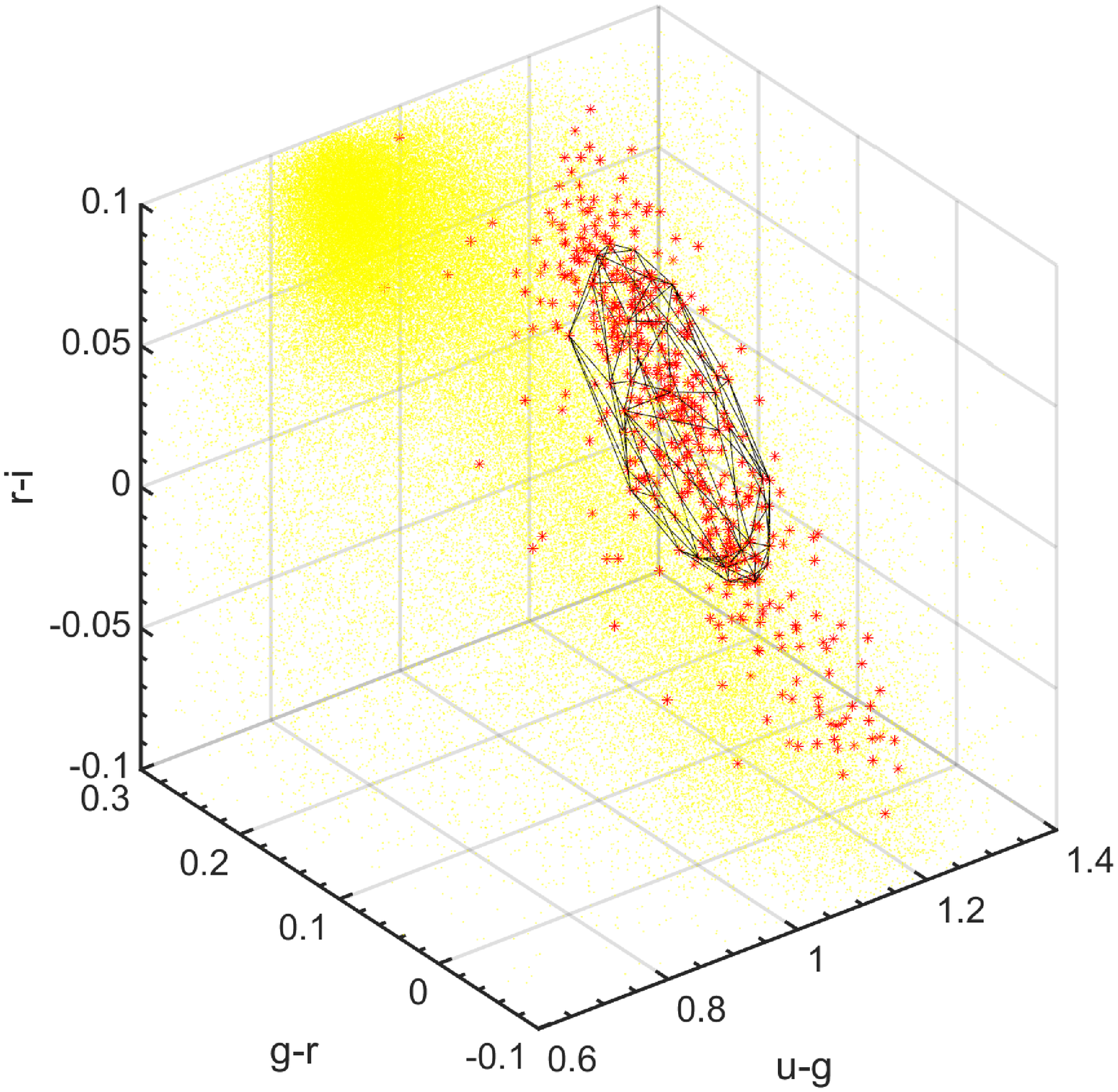}
\caption{Convex hull built in the $u-g, g-r, r-i$ color space at 53\% completeness\label{fig:Fig4}. The red asterisks represent RR Lyrae stars, and the yellow dots represent other stars.}
\label{fig4}
\end{figure}

\begin{figure}
\centering
\includegraphics[bb=288 18 940 660,width=15cm]{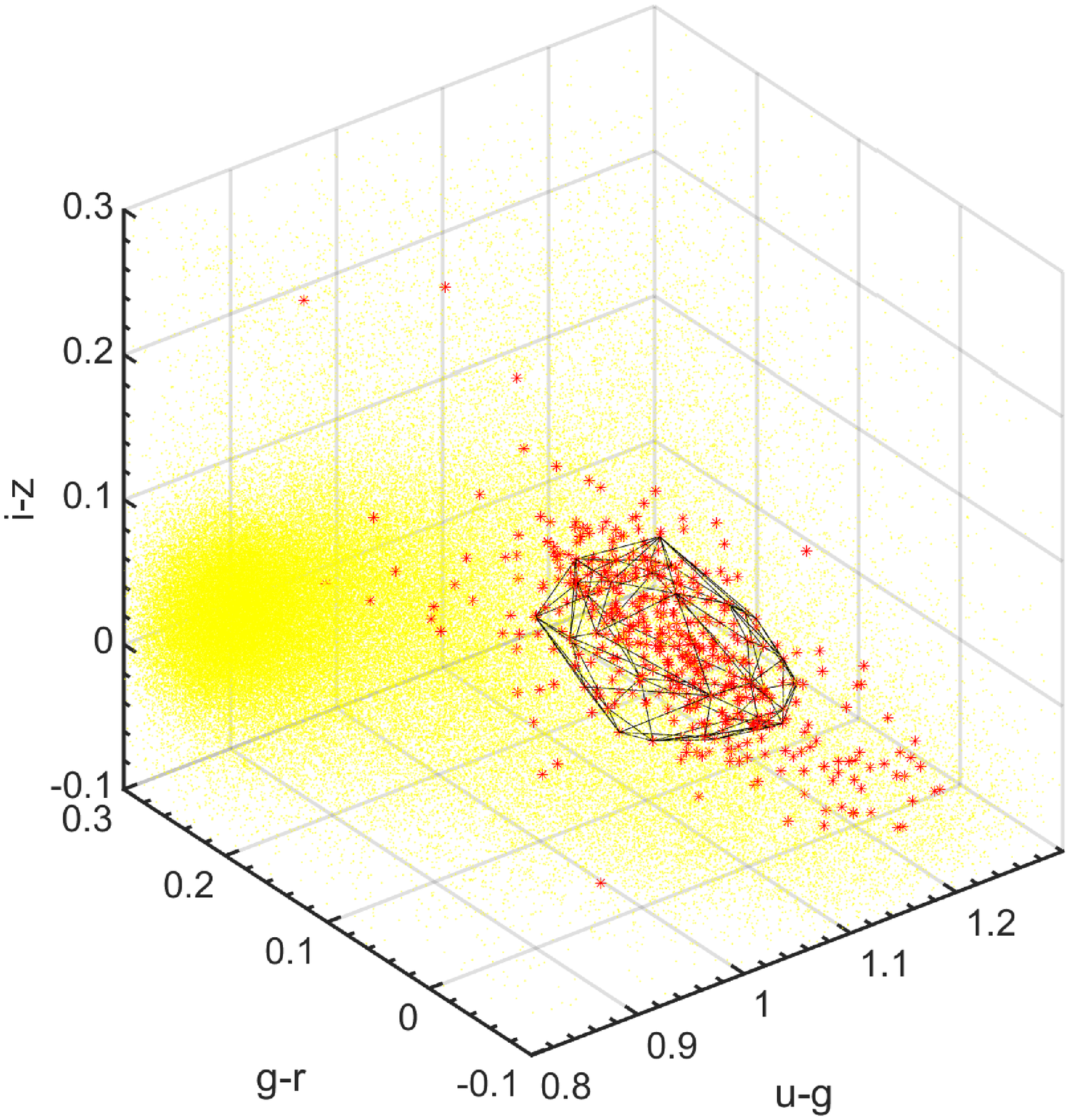}
\caption{Convex hull built in the $u-g, g-r, i-z$ color space at 53\% completeness\label{fig:Fig5}. The red asterisks represent RR Lyrae stars, and the yellow dots represent other stars.}
\label{fig5}
\end{figure}

\begin{figure}
\centering
\includegraphics[bb=288 18 940 660,width=15cm]{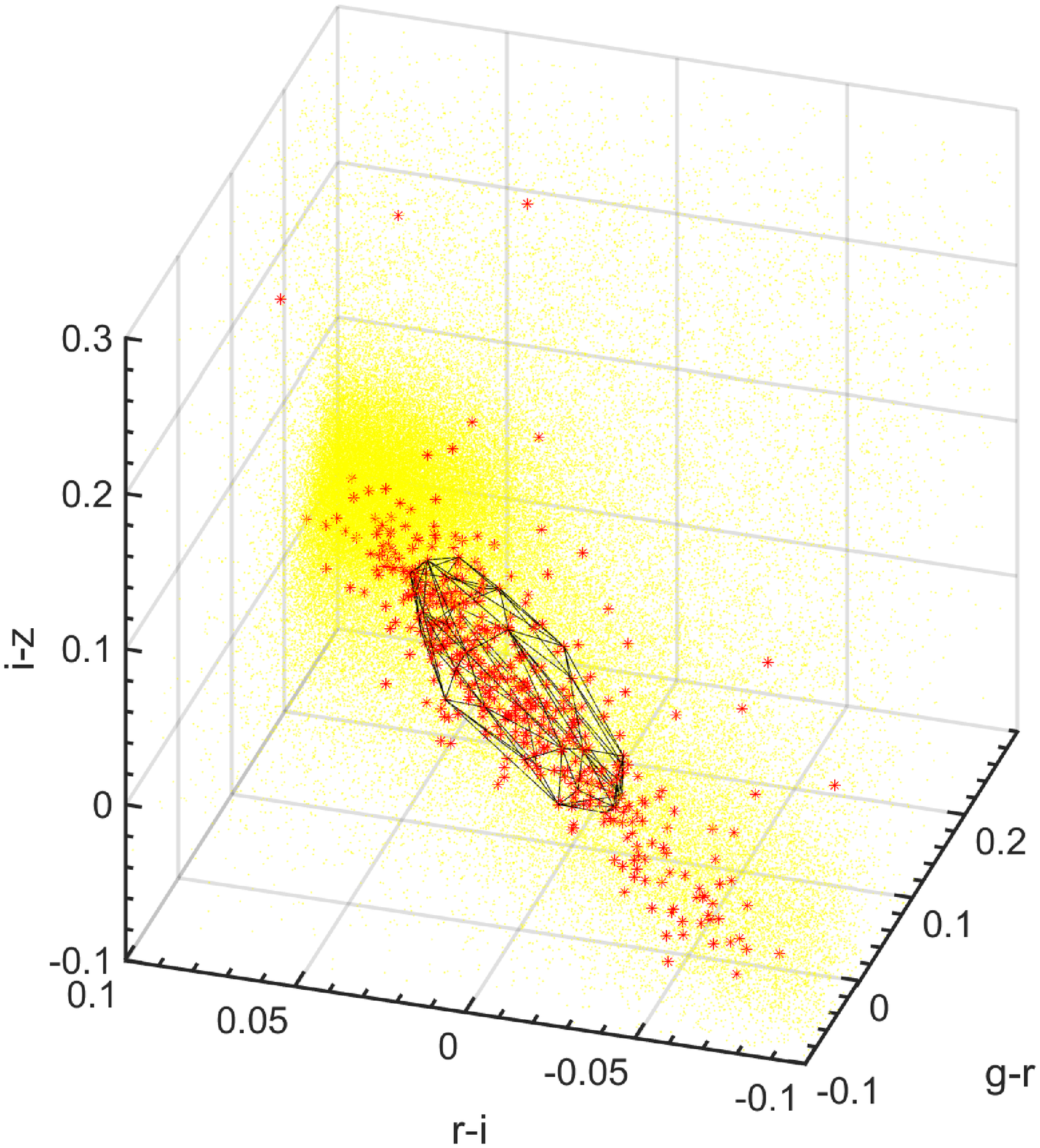}
\caption{Convex hull built in the $g-r, r-i, i-z$ color space at 53\% completeness\label{fig:Fig6}. The red asterisks represent RR Lyrae stars, and the yellow dots represent other stars.}
\label{fig6}
\end{figure}

\begin{figure}[!htb]
\centering
\includegraphics[bb=288 18 940 660,width=15cm]{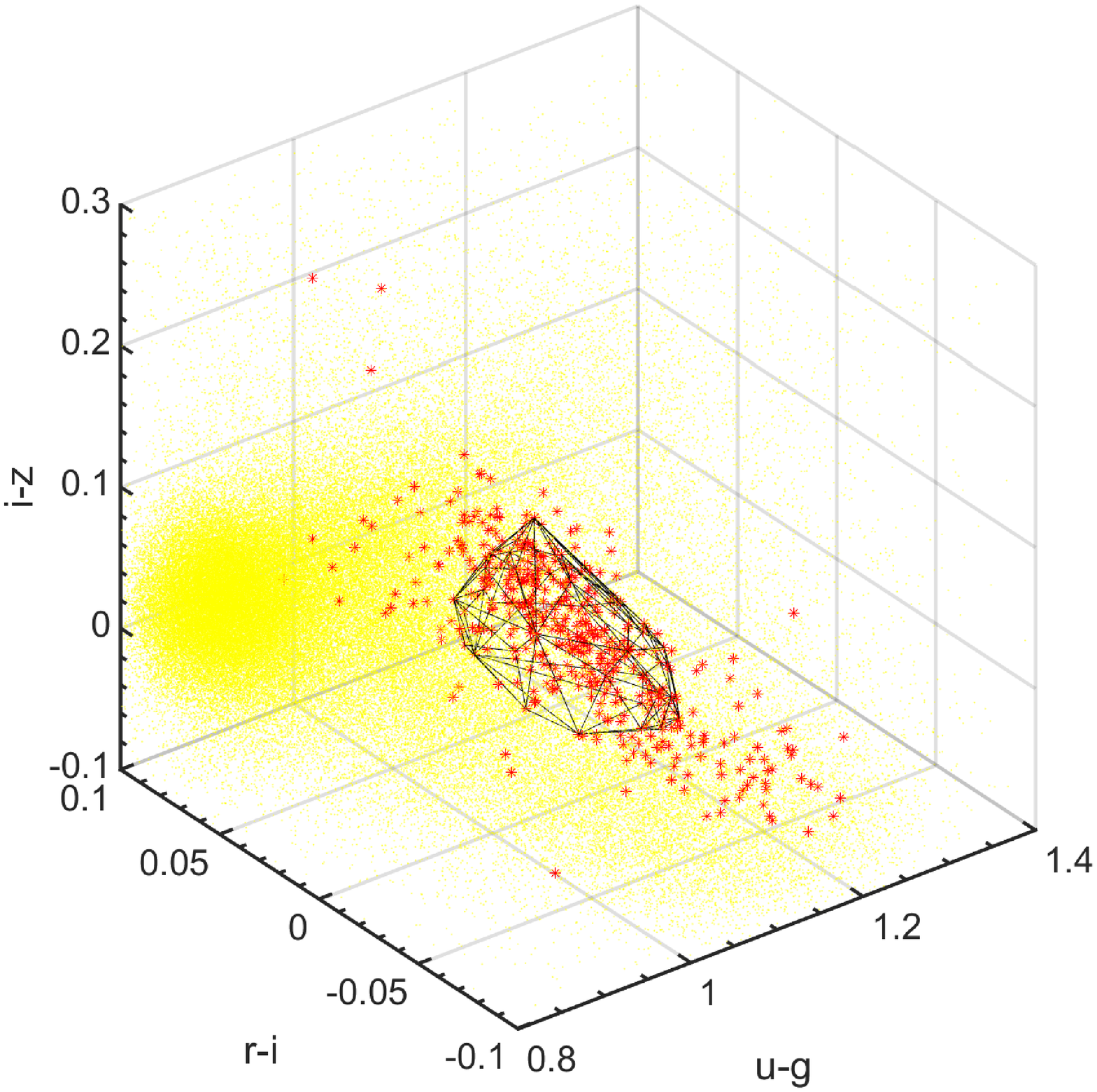}
\caption{Convex hull built in the $u-g, r-i, i-z$ color space at 53\% completeness\label{fig:Fig7}. The red asterisks represent RR Lyrae stars, and the yellow dots represent other stars.}
\label{fig7}
\end{figure}

\begin{figure}
	\centering
	\includegraphics[bb=288 18 940 660,width=15cm]{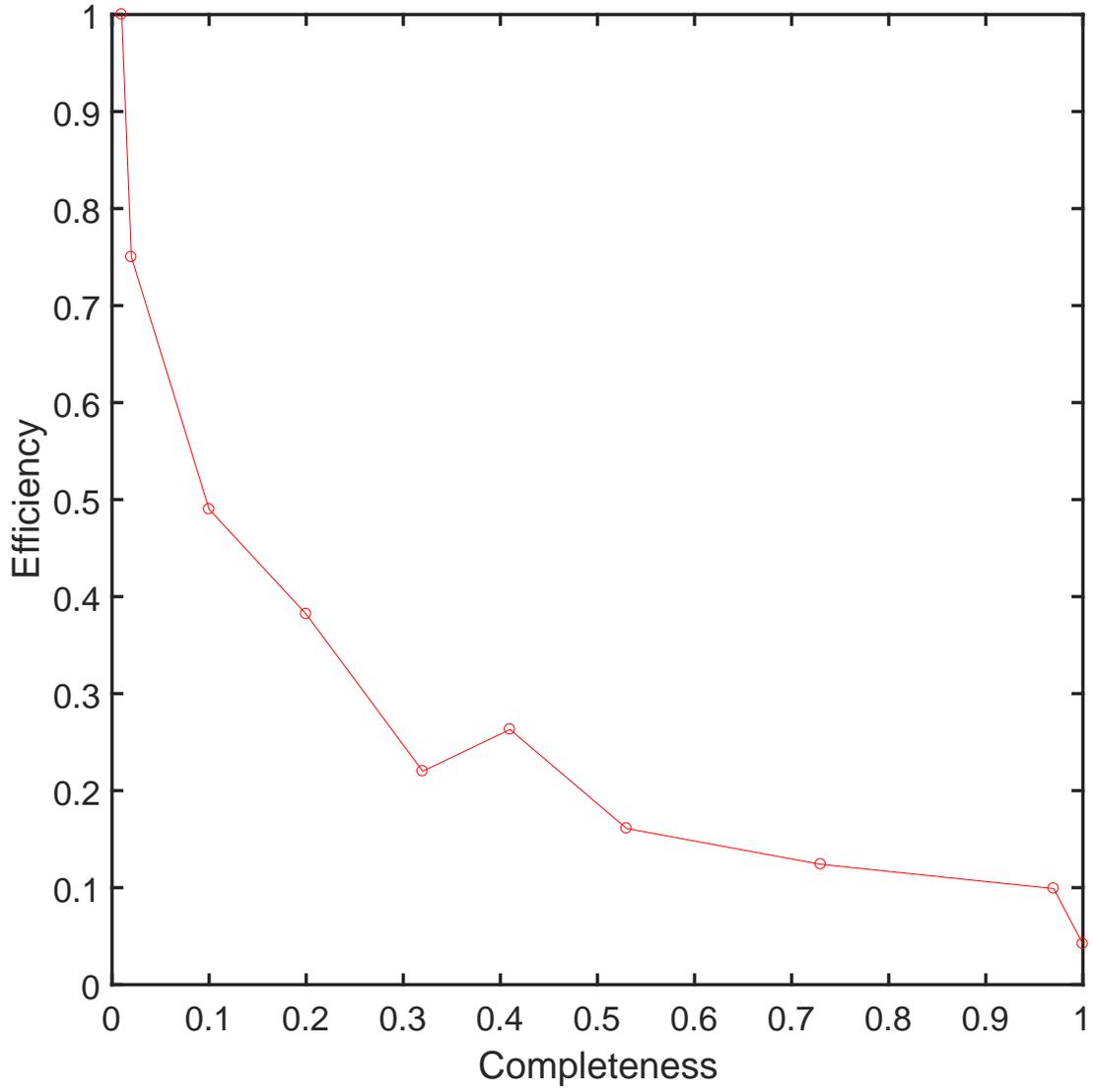}
	\caption{Trade-off between efficiency and completeness with the input pattern ($u-g, g-r, i-z$). \label{fig:Fig8}}
	\label{fig8}
\end{figure}

In 2d and 3d spaces, we build convex hulls with different input patterns based on all 511,201 sources, i.e. ($u-g, g-r$), ($g-r, r-i$), ($r-i, i-z$), ($u-g, g-r, r-i$), ($g-r, r-i, i-z$), ($u-g, g-r, i-z$), ($u-g, r-i, i-z$). For each input pattern, the efficiency of selecting RR Lyrae stars is computed and they are 1.0\%, 0.6\%, 0.5\%, 3.4\%, 2.1\%, 4.2\%, 1.8\% and 1.7\% with 100\% completeness, respectively, as shown in Table \ref{tab:result}. What is more, we compare the performance of convex hull with that of the color cut method in \citet{iz03}. For convex hull algorithm, the efficiency (1.0\%) is the best with the input pattern of ($u-g, g-r$) in a 2-d space, while the efficiency (4.2\%) is the best with the input pattern ($u-g, g-r, i-z$) in a 3-d space. The efficiency of color cuts is 1.7\% when we adopt the same color cut method in \citet{iz03}. As indicated in Table \ref{tab:result}, the performance of \citet{iz03} is higher than the convex hull in a 2-d space, while in a 3-d space, the performances with convex hull algorithm are all better than it. As convex hull algorithm with the best input pattern ($u-g, g-r, i-z$), the efficiency can be increased to 9.9\% with a completeness of 97\% and to 16.1\% with 53\% completeness by removing some outliers. In terms of 53\% completeness, for two-dimensional input patterns, the convex hulls for RR Lyrae stars are shown in Figures~\ref{fig:Fig1}-\ref{fig:Fig3}; for three-dimensional input patterns see Figures~\ref{fig:Fig4}-\ref{fig:Fig7}. Perhaps the higher the dimensions, the better performance the convex hull algorithm has. Due to the computation difficulty of this algorithm in the much higher dimensional spaces, the results with more than three input features are not computed here. Convex hulls in the figures contain more RR Lyrae stars and keep other types of stars as little as possible to reduce the contamination. If further removing much more outliers, the efficiency improves while the completeness decreases. The relation between efficiency and completeness is indicated in Table~\ref{tab:comp5} and Figure~\ref{fig:Fig8}. As shown in Table~\ref{tab:comp5}, if including all sparse RR Lyrae stars, the efficiency reduces to the minimum 4.2\% while the completeness reaches the highest 100\%; the efficiency amounts to 100\% but the completeness is equal to 1\%. When applying the criteria with higher completeness to select RR Lyrae star candidates, more other kinds of stars will be contained. In general, astronomers are inclined to choose high efficiency at the expense of completeness. Therefore we should compromise efficiency and completeness with caution in the real applications.

In a higher dimensional space, cost-sensitive learning is used. Note that undersampling is a wrapper-based approach that makes any learning algorithm cost-sensitive, while the cost-sensitivity is embedded in the cost-sensitive learning algorithm \citep{wg07}. For our case, the number of the majority is large enough while the minority is too small. It is necessary to reduce the size of the majority class and keep different classes balanced. Therefore undersampling is combined with the cost-sensitive learning to train our sample. Undersampling downsamples the majority class randomly to reduce the ratio of the majority to the minority. We use the cross-matched data of SDSS and GALEX photometric data to single out RR Lyrae stars. The number of cross-matched sources is 169,719. They are randomly split into two parts, whose number ratio is 7:3. From the first part, the training set is generated according to the ratios of RR Lyrae stars to the others, which are 1:10, 1:50, 1:100, 1:150, 1:200, 1:250 and 1:300, respectively. The second part is set as the verification set. The detailed workflow is indicated in Figure \ref{fig:Fig9}. For different algorithms, the imbalance weighting parameters of cost-sensitive learning need be adjusted to the optimal in order to compare the performance. The weight for the negative class is $c$, whose value goes from 0.1 to 1 and its step is 0.1. For the cost-sensitive SVM, the 10-fold cross-validation is used to select the optimal $\gamma$ and $C$ parameters. The $\gamma$ parameter ranges from $2^{-5}$ to 1, and the $C$ parameter varies from $2^{-5}$ to $2^{5}$. The cost-sensitive Random Forest also uses the 10-fold cross-validation, which sets the $n$ estimators from 10 to 100. Like \citet{rc14}, the training set for Fast Boxes is separated into 3 folds to select the $K$ and expansion parameters, whose values are 0, 5, 10, $\cdots$, 200. For Fast Boxes, we apply 3-fold cross-validation instead of 10-fold considering the complex computing of Fast Boxes.

To compare the performance of Fast Boxes with SVM and Random Forest better, the input patterns ($u-g, g-r, r-i, i-z$), ($u-g, g-r, r-i, i-z, r$), ($NUV-u, u-g, g-r, r-i, i-z$) and ($NUV-u, u-g, g-r, r-i, i-z, r$) are adopted. For different patterns, we get different results for every algorithm. Tables \ref{tab:comp1}-\ref{tab:comp4} show the performances in terms of mean AUH values and their standard deviations. For each input pattern of SVM and Random Forest with cost-sensitivity, when the ratio of RR Lyrae stars to other types becomes larger, their AUHs become smaller and the performances become worse. But Fast Boxes have a stable performance. All the results show that Fast Boxes are a good performer for highly skewed data. Comparing the AUH of Fast Boxes with different input patterns, that of ($u-g, g-r, r-i, i-z$) is better than that of ($u-g, g-r, r-i, i-z, r$) when only considering optical information; ($NUV-u, u-g, g-r, r-i, i-z, r$) is the best input pattern as given optical and ultraviolet information. They also indicate that added information from GALEX is helpful for identifying RR Lyrae stars. Figure \ref{fig:Fig10} shows the ROC Curve of Fast Boxes with the input pattern ($NUV-u, u-g, g-r, r-i, i-z, r$). In a ROC curve, the true positive rate (Sensitivity) is plotted as a function of the false positive rate (100-Specificity) for different cut-off points. Each red circle on the ROC curve represents a sensitivity/specificity pair corresponding to a particular decision threshold. A test with perfect discrimination (no overlap in the two distributions) has a ROC curve that passes through the upper left corner (100\% sensitivity, 100\% specificity). Therefore the closer the ROC curve is to the upper left corner, the higher the overall accuracy of the test is \citep{ZC93}.

\section{Conclusion} \label{sec:conclusions}

Learning from imbalanced data is still a hot and challenging issue in machine learning, and always faced in reality, such as image recognition, pointed source classification, distinguishing supernovas from large surveys, detecting rare objects or phenomena in astronomy. Since a classifier is aimed to obtain the highest classification accuracy, the classifier is constructed with highly-imbalanced class distribution data set and predicts an unknown sample as the majority class much more frequently than the minority class. By means of the convex hull algorithm, different convex hulls for RR Lyrae stars are created. Automatically and completely, three-dimensional convex hull selects RR Lyrae stars efficiently comparing with the previous studies. In high-dimensional spaces, machine learning algorithms are applied. In terms of highly-skewed data, we adopt undersampling and cost-sensitive learning. Cost-sensitive SVM and cost-sensitive Random Forest are compared to Fast Boxes with different input patterns to separate the RR Lyrae stars from the other types of stars. For our case, Fast Boxes show superiority to the other two machine learning algorithms and are the best classifier for the imbalanced problem. Obviously, for highly skewed data, Fast Boxes is a good performer. \textbf{Here we only use the single-epoch colors and magnitudes.} When multi-epoch data and light-curves are available, we may make use of the varying properties of RR Lyrae as much as possible. Thus a classifier based on Fast Boxes with multi-epoch data and light-curves will be more efficient than that only with single-epoch colors and magnitudes. For example, some large surveys, like PS1 and LSST, can provide more multi-epoch, asynchronous photometric data and the deeper magnitudes of celestial objects, which can be thought of as good data testbeds for various algorithms. As the amount of data increases, the training sample is larger and more complete. Therefore the performance of a classifier further improves and the classifier may be used for selecting much more RR Lyrae stars candidates. In addition, other efficient algorithms dealing with highly-skewed data will be explored in the astronomical applications. Moreover we will select the other special objects from large samples with Fast Boxes.

\section*{Acknowledgements}
We are very grateful to the referee's important comments and suggestions.
This paper is funded by 973 Program 2014CB845700
and the National Natural Science Foundation of China under grant
No.U1731109. We acknowledgment SDSS and GALEX databases. Funding for the Sloan Digital Sky Survey IV has been provided by the Alfred P. Sloan Foundation, the U.S. Department of Energy Office of Science, and the Participating Institutions. SDSS-IV acknowledges support and resources from the Center for High-Performance Computing at the University of Utah. The SDSS web site is www.sdss.org. SDSS-IV is managed by the Astrophysical Research Consortium for the Participating Institutions of the SDSS Collaboration including the Brazilian Participation Group, the Carnegie Institution for Science, Carnegie Mellon University, the Chilean Participation Group, the French Participation Group, Harvard-Smithsonian Center for Astrophysics, Instituto de Astrof\'isica de Canarias, The Johns Hopkins University, Kavli Institute for the Physics and Mathematics of the Universe (IPMU) /University of Tokyo, Lawrence Berkeley National Laboratory, Leibniz Institut f\"ur Astrophysik Potsdam (AIP), Max-Planck-Institut f\"ur Astronomie (MPIA Heidelberg), Max-Planck-Institut f\"ur Astrophysik (MPA Garching), Max-Planck-Institut f\"ur Extraterrestrische Physik (MPE), National Astronomical Observatories of China, New Mexico State University, New York University, University of Notre Dame, Observat\'ario Nacional / MCTI, The Ohio State University, Pennsylvania State University, Shanghai Astronomical Observatory, United Kingdom Participation Group, Universidad Nacional Aut\'onoma de M\'exico, University of Arizona, University of Colorado Boulder, University of Oxford, University of Portsmouth, University of Utah, University of Virginia, University of Washington, University of Wisconsin, Vanderbilt University, and Yale University.

\clearpage

\begin{figure}
\centering
\includegraphics[bb=39 8 760 785,width=12cm,clip]{fig9.eps}
\caption{The detailed workflow\label{fig:Fig9}.}
\label{fig9}
\end{figure}

\begin{figure}
\centering
\includegraphics[bb=288 18 940 660,width=15cm]{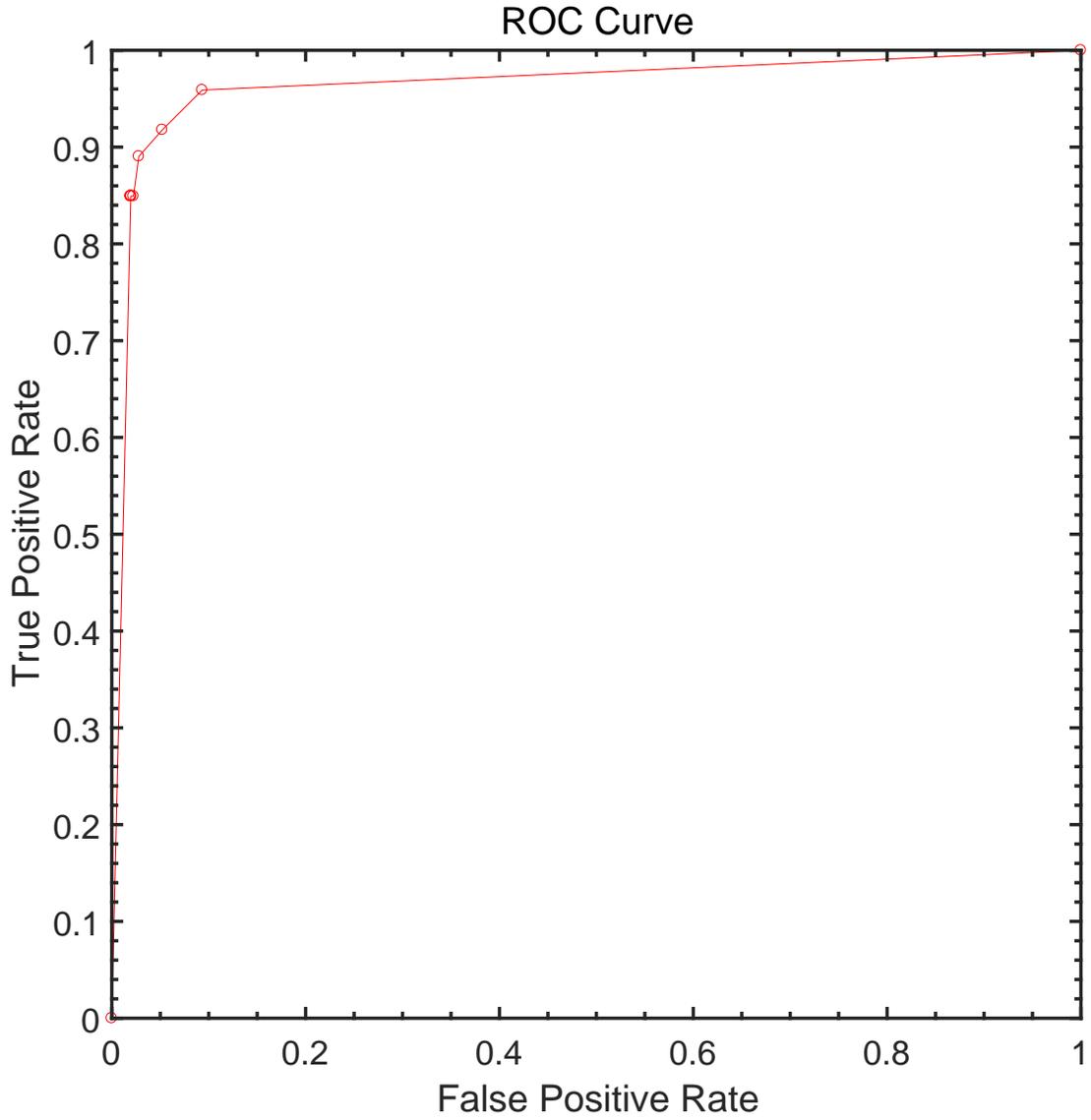}
\caption{The ROC Curve of Fast Boxes with the input pattern ($NUV-u, u-g, g-r, r-i, i-z, r$). Each red circle on the ROC curve represents a sensitivity/specificity pair corresponding to a particular decision threshold. \label{fig:Fig10}}
\label{fig10}
\end{figure}

\clearpage

\begin{table}
\begin{center}
\caption[]{Comparison of the efficiency of the convex hull with that of the color cuts of \citet{iz03} at 100\% completeness.\label{tab:result}}
 \begin{tabular}{ll}
 \hline\noalign{\smallskip} \hline\noalign{\smallskip}
 Method & Efficiency(\%)\\
 \hline\noalign{\smallskip}
convex hull with ($u-g$, $g-r$)                & 1.0 \\
convex hull with ($g-r$, $r-i$)                & 0.6 \\
convex hull with ($r-i$, $i-z$)                & 0.5 \\
convex hull with ($u-g$, $g-r$, $r-i$)         & 3.4 \\
convex hull with ($g-r$, $r-i$, $i-z$)         & 2.1 \\
convex hull with ($u-g$, $g-r$, $i-z$)         & 4.2 \\
convex hull with ($u-g$, $r-i$, $i-z$)         & 1.8 \\
color cuts of \citet{iz03}                     & 1.7\\
\noalign{\smallskip}\hline
\end{tabular}
\end{center}
\end{table}

\clearpage

\begin{table}
	\begin{center}
		\caption[]{Completeness and efficiency with the input pattern ($u-g, g-r, i-z$). \label{tab:comp5}}
		\begin{tabular}{llccccccccc}
			\hline\noalign{\smallskip} \hline\noalign{\smallskip}
         Completeness&1.000 &0.970 &0.730 &0.530 &0.410 &0.320 &0.200 &0.100 &0.020 &0.010\\
         Efficiency  &0.042 &0.099 &0.124 &0.161 &0.263 &0.220 &0.382 &0.490 &0.750 &1.000\\
			\noalign{\smallskip}\hline
		\end{tabular}
	\end{center}
\end{table}

\clearpage

\begin{table}
\begin{center}
\caption[]{Comparison of AUH of Fast Boxes with those of cost-sensitive SVM and cost-sensitive Random Forest with ($u-g, g-r, r-i, i-z$). \label{tab:comp1}}
 \begin{tabular}{r|ccc}
 \hline\noalign{\smallskip} \hline\noalign{\smallskip}
The ratio of positive sample to negative sample &SVM & Random Forest & Fast Boxes\\
                                      &AUH & AUH           & AUH\\
 \hline\noalign{\smallskip}
1 : 10  & 0.98$\pm$0.00 & 0.96$\pm$0.01 & 0.97$\pm$0.02 \\
1 : 50  & 0.87$\pm$0.02 & 0.86$\pm$0.02 & 0.95$\pm$0.02 \\
1 : 100 & 0.50$\pm$0.00 & 0.81$\pm$0.01 & 0.96$\pm$0.00 \\
1 : 150 & 0.50$\pm$0.00 & 0.74$\pm$0.03 & 0.94$\pm$0.02 \\
1 : 200 & 0.50$\pm$0.00 & 0.69$\pm$0.02 & 0.94$\pm$0.02 \\
1 : 250 & 0.50$\pm$0.00 & 0.65$\pm$0.01 & 0.94$\pm$0.01 \\
1 : 300 & 0.50$\pm$0.00 & 0.61$\pm$0.00 & 0.96$\pm$0.01 \\
\noalign{\smallskip}\hline
\end{tabular}
\end{center}
\end{table}

\clearpage

\begin{table}
\begin{center}
\caption[]{Comparison of AUH of Fast Boxes with those of cost-sensitive SVM and cost-sensitive Random Forest with ($u-g, g-r, r-i, i-z, r$). \label{tab:comp2}}
 \begin{tabular}{r|ccc}
 \hline\noalign{\smallskip} \hline\noalign{\smallskip}
The ratio of positive sample to negative sample &SVM & Random Forest & Fast Boxes\\
                                      &AUH & AUH           & AUH\\
 \hline\noalign{\smallskip}
1 : 10  & 0.98$\pm$0.01 & 0.96$\pm$0.01 & 0.97$\pm$0.01 \\
1 : 50  & 0.87$\pm$0.02 & 0.87$\pm$0.01 & 0.95$\pm$0.03 \\
1 : 100 & 0.51$\pm$0.00 & 0.78$\pm$0.02 & 0.96$\pm$0.00 \\
1 : 150 & 0.51$\pm$0.01 & 0.71$\pm$0.02 & 0.97$\pm$0.01 \\
1 : 200 & 0.50$\pm$0.00 & 0.68$\pm$0.01 & 0.94$\pm$0.00 \\
1 : 250 & 0.50$\pm$0.00 & 0.62$\pm$0.01 & 0.96$\pm$0.00 \\
1 : 300 & 0.50$\pm$0.00 & 0.61$\pm$0.01 & 0.96$\pm$0.00 \\
\noalign{\smallskip}\hline
\end{tabular}
\end{center}
\end{table}

\clearpage

\begin{table}
\begin{center}
\caption[]{Comparison of AUH of Fast Boxes with those of cost-sensitive SVM and cost-sensitive Random Forest with ($NUV-u, u-g, g-r, r-i, i-z$). \label{tab:comp3}}
 \begin{tabular}{r|ccc}
 \hline\noalign{\smallskip} \hline\noalign{\smallskip}
The ratio of positive sample to negative sample & SVM & Random Forest & Fast Boxes\\
                                       & AUH & AUH           &AUH\\
 \hline\noalign{\smallskip}
1 : 10  & 0.99$\pm$0.00 & 0.95$\pm$0.01 & 0.98$\pm$0.00 \\
1 : 50  & 0.85$\pm$0.04 & 0.80$\pm$0.01 & 0.96$\pm$0.01 \\
1 : 100 & 0.65$\pm$0.03 & 0.77$\pm$0.01 & 0.96$\pm$0.00 \\
1 : 150 & 0.50$\pm$0.00 & 0.71$\pm$0.00 & 0.94$\pm$0.01 \\
1 : 200 & 0.50$\pm$0.00 & 0.69$\pm$0.01 & 0.95$\pm$0.02 \\
1 : 250 & 0.50$\pm$0.00 & 0.65$\pm$0.00 & 0.95$\pm$0.01 \\
1 : 300 & 0.50$\pm$0.00 & 0.61$\pm$0.01 & 0.96$\pm$0.00 \\
\noalign{\smallskip}\hline
\end{tabular}
\end{center}
\end{table}

\clearpage

\begin{table}
\begin{center}
\caption[]{Comparison of AUH of Fast Boxes with those of cost-sensitive SVM and cost-sensitive Random Forest with ($NUV-u, u-g, g-r, r-i, i-z, r$). \label{tab:comp4}}
 \begin{tabular}{r|ccc}
 \hline\noalign{\smallskip} \hline\noalign{\smallskip}
The ratio of positive sample to negative sample & SVM & Random Forest &Fast Boxes\\
                                          & AUH & AUH           & AUH\\
 \hline\noalign{\smallskip}
1 : 10  & 0.96$\pm$0.01 & 0.95$\pm$0.01 & 0.97$\pm$0.01\\
1 : 50  & 0.84$\pm$0.03 & 0.85$\pm$0.01 & 0.97$\pm$0.00\\
1 : 100 & 0.51$\pm$0.00 & 0.76$\pm$0.02 & 0.97$\pm$0.06\\
1 : 150 & 0.52$\pm$0.00 & 0.71$\pm$0.00 & 0.96$\pm$0.01\\
1 : 200 & 0.50$\pm$0.01 & 0.64$\pm$0.02 & 0.94$\pm$0.01\\
1 : 250 & 0.51$\pm$0.01 & 0.63$\pm$0.01 & 0.96$\pm$0.01\\
1 : 300 & 0.50$\pm$0.01 & 0.61$\pm$0.01 & 0.96$\pm$0.00\\
\noalign{\smallskip}\hline
\end{tabular}
\end{center}
\end{table}


\begin{thebibliography}{}
\bibitem[Abbas et~al.(2014)]{gc14}Abbas, M., et~al. 2014, \aj, 148, 2782
\bibitem[Armstrong et~al.(2016)]{ad16}Armstrong, D., Kirk, J., Lam, K., et~al. 2016, \mnras, 456, 2260
\bibitem[Bianchi et~al.(2011)]{bh11}Bianchi, L., Herald, J., Efremova, B., Girardi, L., Zabot, A., Marigo, P., Conti, A., Shiao, B. 2011, Ap\&SS, 335, 161
\bibitem[Breiman et~al.(2001)]{br01}Breiman, L. 2001, Machine Learning, 45, 5
\bibitem[Bullock et~al.(2001)]{bu01}Bullock, J. S., Kravtsov, A. V., Weinberg, D. H. 2001, \apj, 539, 517
\bibitem[Chawla et~al.(2002)]{ch02}Chawla, N. V., Bowyer, K. W., Hall, L. O., et~al. 2002, Journal of Artificial Intelligence Research, 16, 321
\bibitem[Davenport et~al.(2014)]{dj14}Davenport, J. R. A., Ivezi\'c, Z., Becker, A. C., et~al. 2014, \mnras, 440, 3430
\bibitem[Elorrieta et~al.(2016)]{ef16}Elorrieta, F., Eyheramendy, S., Jord\'an, A., et~al. 2016, A\&A, 595, A82
\bibitem[Fan (1999)]{fan99}Fan, X. 1999, \aj, 117, 2528
\bibitem[Finlator et~al.(2000)]{fi00}Finlator, K., Ivezi\'c, Z., Fan, X., Strauss, M. A., et~al. 2000, \aj, 32, 2615
\bibitem[Gao et~al.(2009)]{gd09}Gao, D., Zhang, Y., \& Zhao, Y. 2009, RAA, 9, 220
\bibitem[Hernitschek et~al.(2016)]{hn16}Hernitschek, N., Schlafly, E F., \& Sesar, B., et~al. 2016, \apj, 817, 73
\bibitem[Ivezi\'c et~al.(2000)]{iz00}Ivezi\'c, Z., Goldston, J., Finlator, K., et~al. 2000, \aj, 120, 963
\bibitem[Ivezi\'c et~al.(2005)]{iz03}Ivezi\'c, Z., Vivas, A K., Lupton, R H., et~al. 2005, \aj, 129, 1096
\bibitem[Ivezi\'c et~al.(2009)]{iz08}Ivezi\'c, Z., et~al. 2009, American Astronomical Society Meeting Abstracts \#213, Bulletin of American Astronomical Society, 41, 366
\bibitem[Krisciunas (2001)]{kk01}Krisciunas, K. 2001, \pasp , 113, 121
\bibitem[Krisciunas et~al.(1998)]{kk98}Krisciunas, K., Margon, B., Szkody, P. 1998, \pasp, 110, 1342
\bibitem[Peng et~al.(2013)]{pn13}Peng, N., Zhang, Y., Zhao, Y. 2013, Science CHINA, Physics, Mechanics \& Astronomy, 56, 1227
\bibitem[Pedregosa et~al.(2011)]{pe11}Pedregosa, et~al. 2011, JMLR, 12, 2825
\bibitem[Provost \& Fawcett (2001)]{pfft01}Provost, F., Fawcett, T. 2001, Kluwer Academic Publishers, 42, 203
\bibitem[Rudin et~al.(2014)]{rc14}Goh, S. T., \& Rudin, C. 2014, ACM SIGKDD International Conference on Knowledge Discovery and Data Mining, 333
\bibitem[Schlegel et~al.(1998)]{sc98}Schlegel, D. J., Finkbeiner, D. P., Davix, M. 1998,  \apj, 500, 525
\bibitem[Sesar et~al.(2017)] {sb17}Sesar, B., Hernitschek, N., Mitrovi\'c, S., et~al. 2017,  \aj, 153, 204
\bibitem[Sesar et~al.(2007)]{sb07}Sesar, B., Ivezi\'c, Z., Lupton, R. H., et~al. 2007, \aj, 134, 2236
\bibitem[Sklansky (1982)]{sk82}Sklansky, J. 1982, Pattern Recognition Letters, 1, 79
\bibitem[Skrutskie et~al.(2006)]{sk06} Skrutskie, M. F., Cutri, R. M., Stiening, R., et~al. 2006, \aj, 131, 1163
\bibitem[Shrivastava et~al.(2015)]{sh15}Shrivastava, H., Huddar, V., Bhattacharya, S., et~al. 2015, IEEE International Conference, 707
\bibitem[Vapnik (1995)]{vv95}Vapnik, V. 1995, The Nature of Statistical Learning Theory, 2nd edn. Springer-Verlag, New York, 89
\bibitem[Weiss et~al.(2007)]{wg07}Weiss, G., McCarthy, K., \&Zabar, B. 2007, International Conference on Data Mining, 35
\bibitem[Wright et~al.(2010)]{wr10}Wright, E. L., Eisenhardt, P. R. M., Mainzer, A. K., et~al. 2010, \aj, 140, 1868
\bibitem[Yan et~al.(2017)]{ya17}Yan, Q., Xia, S., Meng, F. 2017, arXiv:1702.01504
\bibitem[Yin et~al.(2014)]{yi14}Yin, H., Yuping, H., Information, S. O. 2014, Engineering Journal of Wuhan University
\bibitem[York et~al.(2000)]{york00}York, D. G., et~al. 2000, \aj, 120, 1579
\bibitem[Zweig \& Campbell(1993)]{ZC93}Zweig, M. H., Campbell, G. 1993, Clinical Chemistry, 39, 561
\end{thebibliography}
\end{document}